\documentclass{kluwer}   
\usepackage{epsfig}

\begin{document}
\begin{article}

\begin{opening}

\title{The dusty SF history of high-z galaxies, modelling tools and future prospects}

\author{Gian Luigi \surname{Granato} \email{granato@pd.astro.it}}
\institute{Osservatorio Astronomico di Padova, Padova, Italy and SISSA, Trieste, Italy}
\author{Laura \surname{Silva} \email{silva@ts.astro.it}}
\institute{Osservatorio Astronomico di Trieste and SISSA, Trieste, Italy}

\begin{abstract} We summarize recent advances in the determination of the cosmic
history of star formation and other properties of high-z
galaxies, and the relevance of this information in our
understanding of the formation of structures. We emphasize the
importance of dust reprocessing in the high--z universe, as
demonstrated in particular by IR and sub-mm data. This demand a
panchromatic approach to observations and suitable modelling
tools. We spend also some words on expectations from future
instruments.
\end{abstract}

\end{opening}

\section{Introduction}

In the last few years, a huge number of studies of the high-z
universe have been devoted to the determination of the cosmic
history of star formation $SFR(z)$. The main motivation for these
efforts is that baryons are the only observational tracers of the
evolution of large scale structures, which are believed to be
driven by the gravitational collapse of dark matter (DM). The
determination of $SFR(z)$ can in principle help to discriminate
between different proposed scenarios for the formation and
evolution of galaxies.

A fundamental issue connected to the determination of $SFR(z)$ is
the fraction of light produced by stars which has been reprocessed
by dust into IR photons. We know that in the local universe
surveyed by IRAS this fraction is $\sim 30\%$. Though this is a
significative percentage, it is not dramatic, in the sense that if
the same were true also at high--z, optical--UV observations would
determine $SFR(z)$ with a small uncertainty. But IRAS
observations demonstrated also that dust reprocessing in
local galaxies is a strong increasing function of their star
formation activity, and can't be reliably determined by UV and
optical data alone. This is vividly illustrated for instance by
figure 2 of Sanders \& Mirabel (1996), which shows that
infrared-selected galaxies range over 3 orders of magnitude in
$L_{IR}$, while the optical luminosity changes only by a factor of
3-4, with minor differences in  the shape of the optical-UV SED.
A more model dependent example of this is shown in Fig.~1. The
possibility to measure the amount of dust hidden SF from the
optical-UV SED has been discussed quite a lot, for instance using
the Meuer's correlation between the UV spectral index and the
ratio $L_{UV}/L_{IR}$. This relationship is found to be quite
narrow for some classes of starburst galaxies, however it has
recently been shown that it does
not hold for the most luminous starburst galaxies at low redshift
(Meuer et al.\ 1999; Meuer et al.\ 2000; see also Panuzzo et al, these
proceedings).

It is therefore quite natural to suspect that in the younger, more
active universe an higher fraction of star luminosity was
reprocessed by dust, and that in this situation the global SF
activity can be reliably measured only by means of FIR and sub-mm
data.

\begin{figure} 
\centerline{\epsfig{file=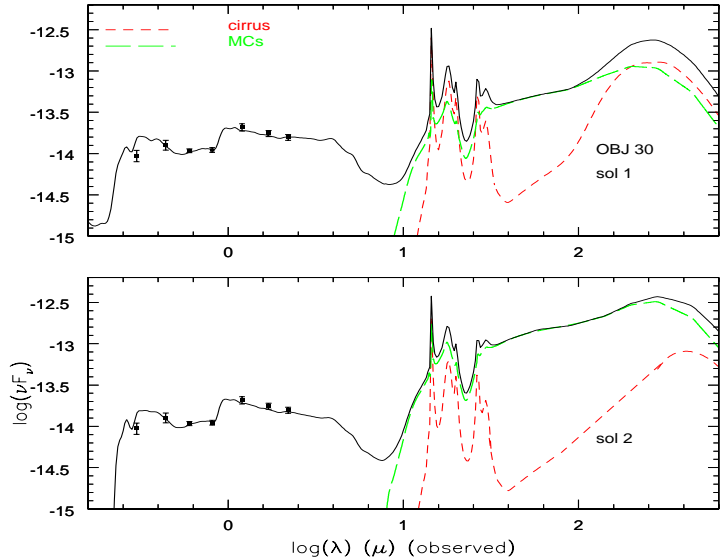,width=10truecm,height=8truecm}}
\caption{The observed optical-NIR SED of a late--type galaxy in
HDF-N at z=1.3 is reproduced with plausible models differing by
a factor 3 in SFR. Adapted from Rodighiero et al.\ (2000).}
\end{figure}

\section{The dusty SF history of galaxies}

Indeed, several pieces of evidence in the paste few years have
shown that {\it most of the SF in the high--z universe is dust
obscured or dimmed to a substantial  degree:}

\begin{itemize}

\item[(1)]
The discovery of a cosmic far-IR/sub-mm background by the COBE
satellite (Puget et al.\ 1996; Fixsen et al.\ 1998; Hauser et al.\ 1998), 
whose energy density, which is at least a factor 2 larger
than the optical-UV one, indicates that a large fraction of the
energy radiated by stars over the history of the universe has
been reprocessed by dust.

\item[(2)]
The SCUBA detection of a population of bright $850 \mu$m sources.
Surveys at around 1 mm are very effective in discovering high-z
dust enshrouded star formation, due to the steep increase of the
expected SED of dusty star forming galaxies shortward this
wavelength (see Silva et al.\ in these proceedings). 
Indeed, the resulting strong positive K
correction is able to counterbalance the cosmological dimming
with increasing z for $1 \lesssim z \lesssim 8$ (details depend
of course on the precise $\lambda$ of the survey and on the cosmological
parameters). As a result, the expected mm flux of a given
luminosity source remains almost constant in this redshift range.
Since the comoving cosmic volume element increases up to $z\sim
2$, the conclusion is that mm or sub-mm surveys are strongly
biased in favor of dusty high redshift objects.

In the last few years this fact has been exploited in particular
by surveys at $850 \mu$m performed with the SCUBA camera at JCMT.
These surveys led to the discovery of a population of sub-mm
sources at high redshift ($z\gtrsim 1.5$), whose luminosities, if
powered by star formation in dust-enshrouded galaxies,
imply very large star formation rates ($\gtrsim 10^2
M_{\odot}\mbox{yr}^{-1}$), and a total star formation density
probably greater than that inferred from the UV luminosities of
the Lyman-break galaxies (Smail et al.\ 1997; Hughes et al.\ 1998).
Also, this bright population is originating a significant
fraction of the IR background (eg.\ Blain 2000)

\item[(3)]
The ISO detection of a population of strong IR sources; 15 $\mu$m
ISOCAM (Oliver et al 1997) and 175 $\mu$m ISOPHOT surveys (Kawara
et al 1998; Puget et al 1999) show a population of actively star
forming galaxies at $0.4 < z < 1.3$, mostly disk/interacting
galaxies with K typical of a L* galaxy. This population boosts
the cosmic star formation density by a factor $\sim 3$ with
respect to that estimated in the optical from the CFRS in the
same redshift range.

\end{itemize}

For (1) and (2), there is the caveat that the contribution from
dust-enshrouded AGNs to the sub-mm counts and background is
currently uncertain, but probably the AGNs do not dominate (e.g.\
Granato, Danese \& Franceschini 1997).

A summary of the present status of the determination of $SFR(z)$
is given for instance by figure 17 of Genzel \& Cesarsky (2000).
The main point is that while a few years ago it was claimed,
based on optical observations, that this function had a peak at
$z\simeq 1$ and declined at higher redshifts, it is now clear
instead that $SFR(z)$ steeply increases from $z=0$ to $z=1$,
but than stays essentially flat to at least $z\simeq 4$.

\section{Interpretative tools}

These observations should be framed in the hierarchical structure
formation paradigm. In this scenario, structures result from the
gravitational amplification of small, primordial density
fluctuations, possibly quantum ripples boosted to macroscopic
scales by inflation. The subsequent formation and merging of 
DM haloes is driven by gravitation and fully determined
by the initial density fluctuations and by the cosmology.

While the simulation of the behaviour of the DM component of the
universe is a relatively simple task, a much more difficult
problem is to predict the evolution of the baryonic component, 
subject to a much more complex physics.
As a matter of fact, to date it is not possible to follow by
direct numerical simulations these processes with the dynamical
range relevant for galaxy formation. The most employed technique
is instead that of semi-analytical models, in which simplified
analytical descriptions of baryon processes (e.g.\ gas cooling
and collapse, star formation, supernovae feedback and galaxy
merging) are adopted (see Cole et al.\ 2000). The predicted SF
histories are then combined with stellar population models
including dust reprocessing, in order to predict galaxy
observational properties.

\begin{figure} 
\centerline{\epsfig{file=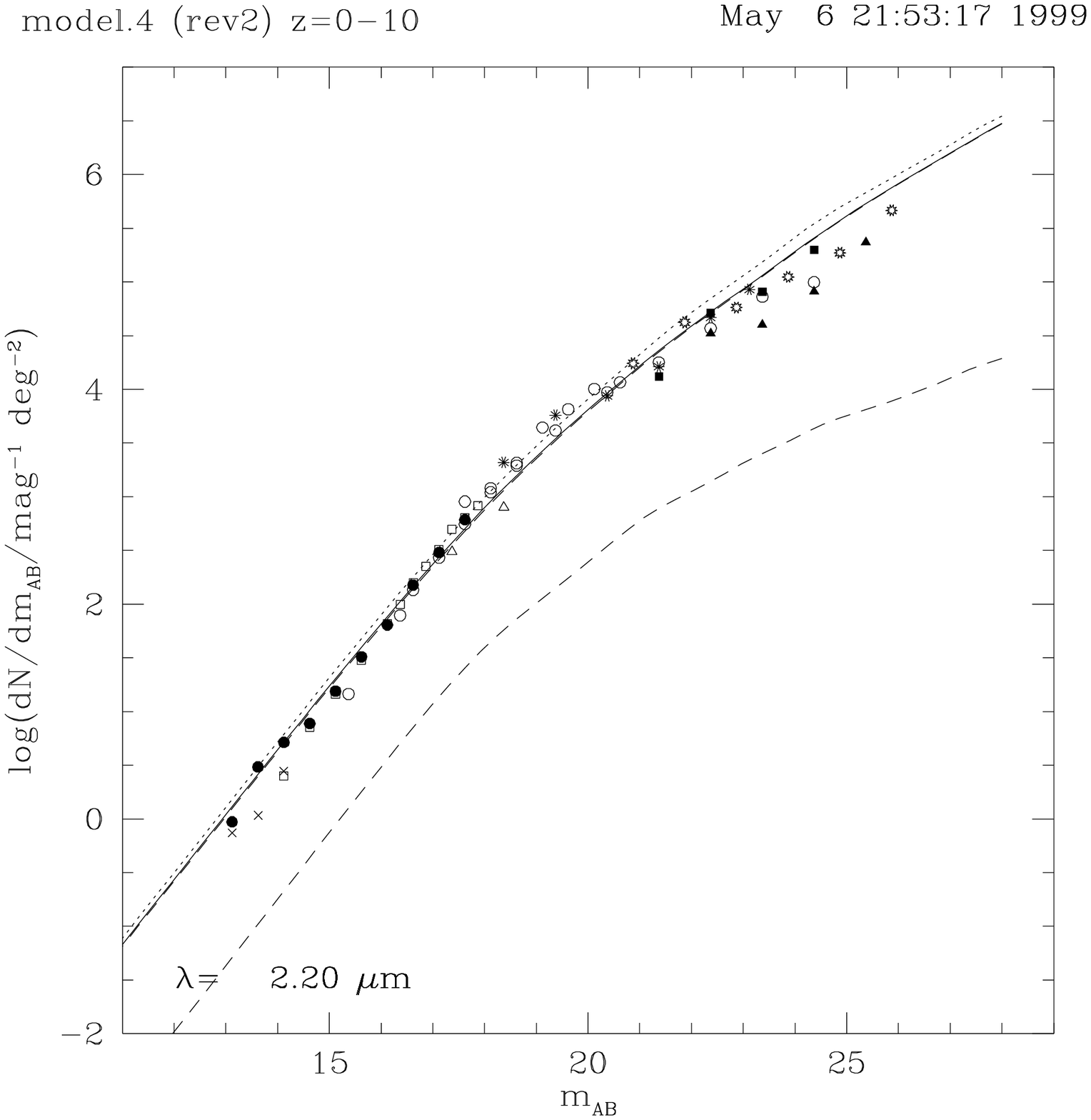,height=6truecm,width=6truecm}
 \epsfig{file=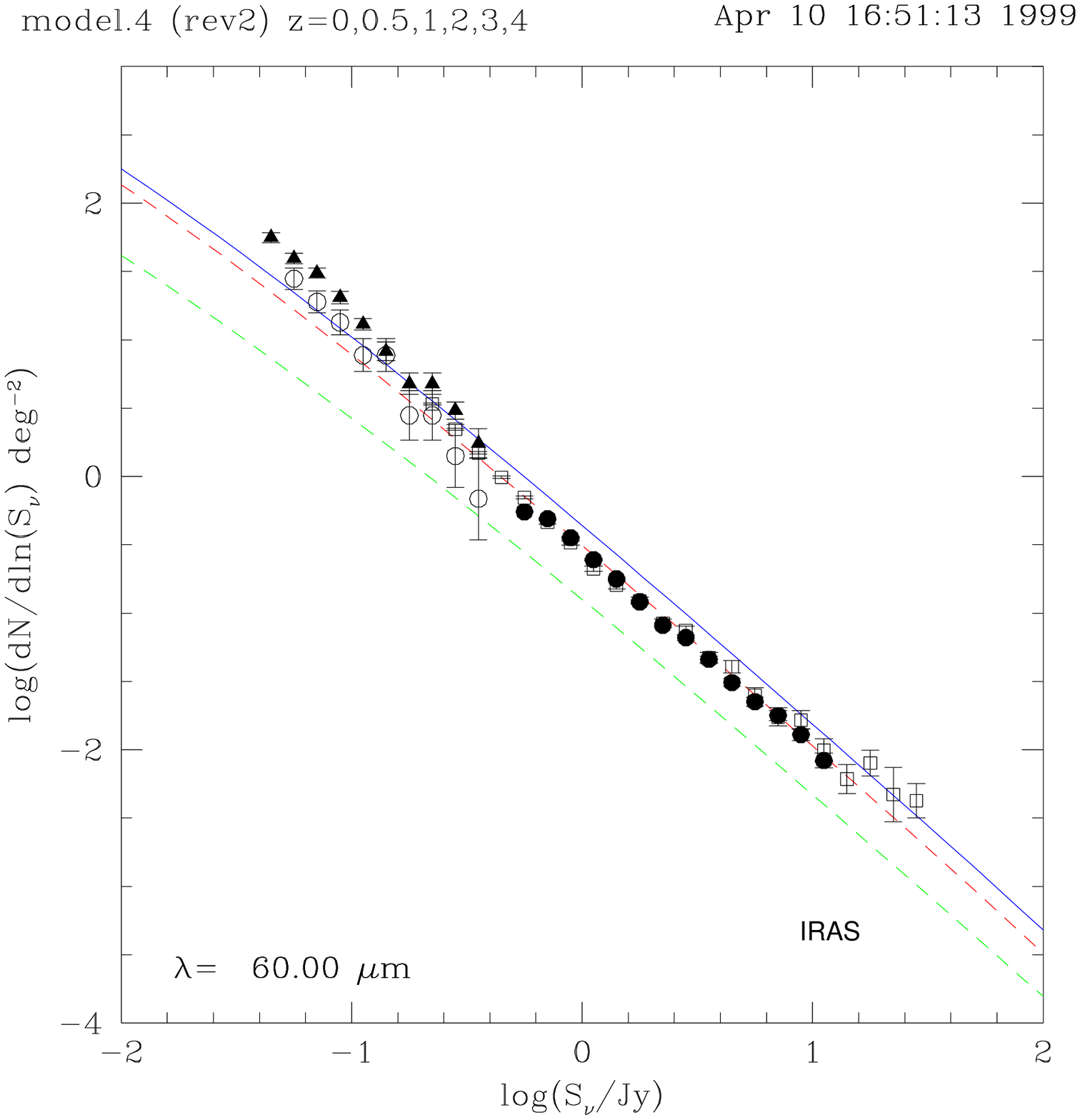,height=6truecm,width=6truecm} } 
\centerline{ \epsfig{file=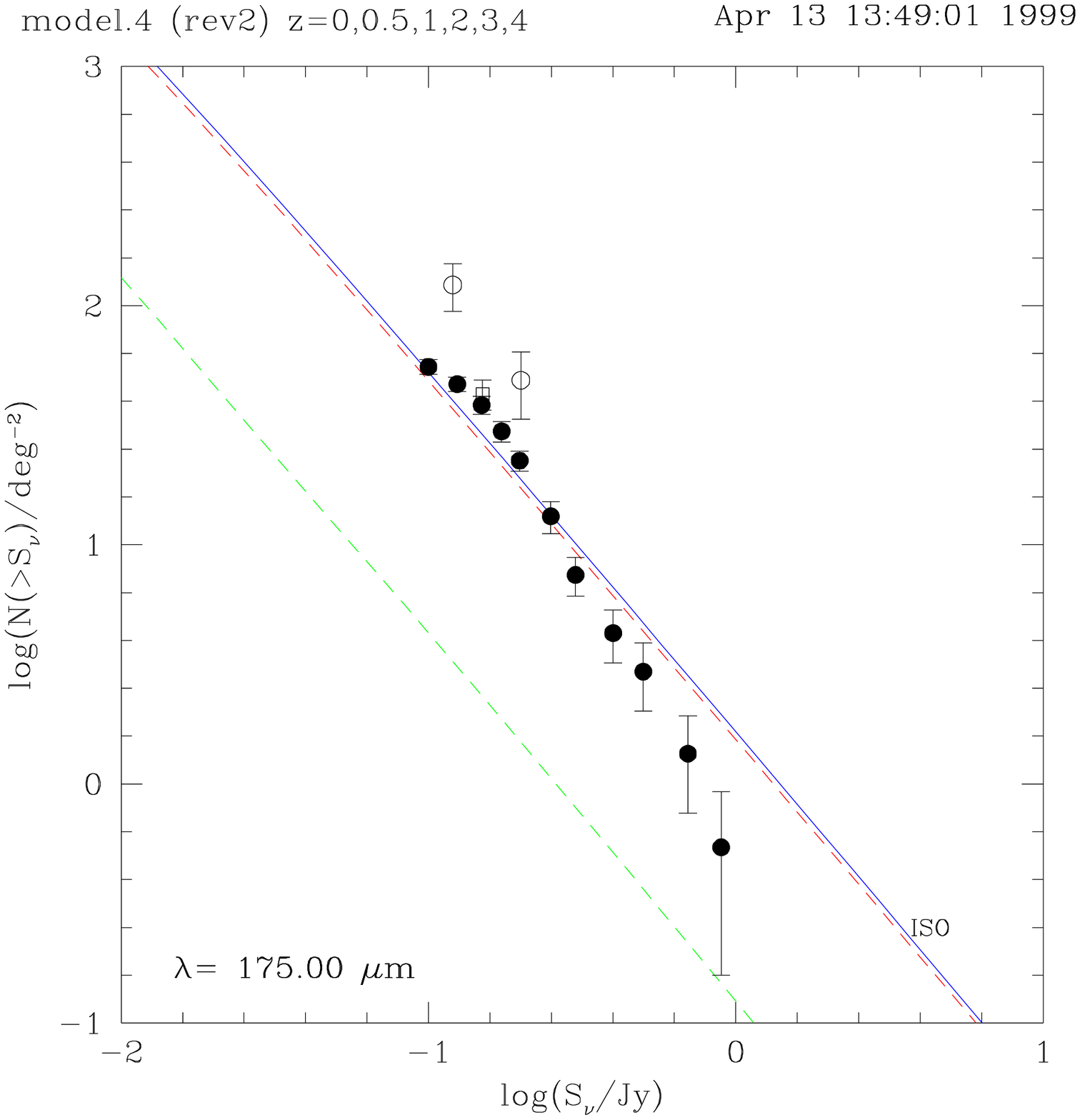,height=6truecm,width=6truecm}  
 \epsfig{file=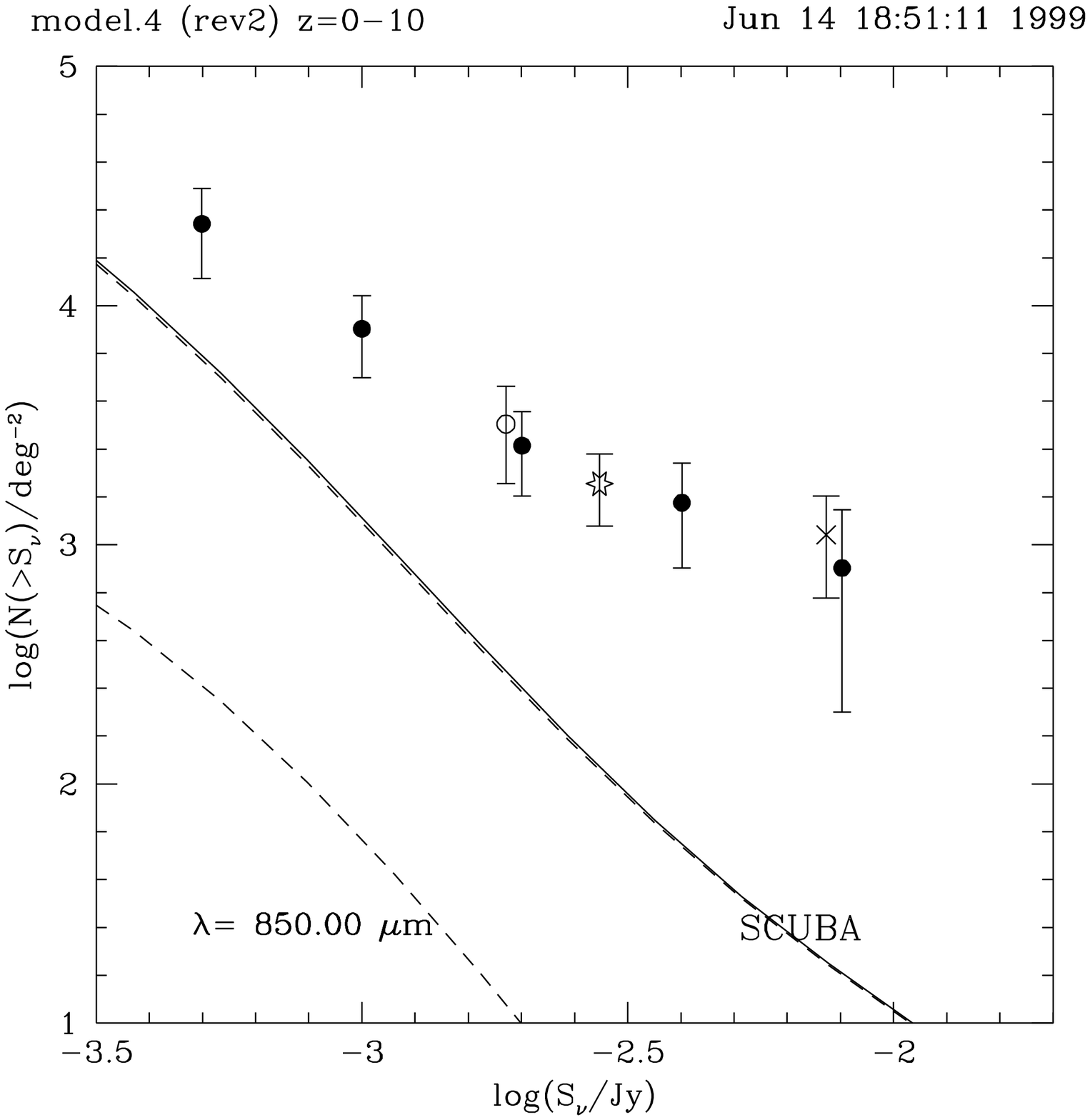,height=6truecm,width=6truecm} }
\caption{Number counts at 2.2, 60, 175 and 850 $\mu$m. 
Standard semi-analytical models fail to reproduce the
$850\mu$m SCUBA counts. Some substantial revision of the
prescriptions used for star formation at high redshift seems
unavoidable (Silva 1999; see also Lacey et al 2001, in preparation)}
\end{figure}

Present semi-analytical models compare favorably with most
observations, including many IR constraints (Granato et al 2000
and references therein; Silva 1999) but are seriously challenged
by sub-mm counts, which turn out to be underpredicted by about
one order of magnitude (Fig.~2). This, coupled with the fact that
these models are instead able to reproduce the observed
background from the FUV to the sub-mm, leads to the conclusion that
the main problem is that they spread the total SF activity in too many
but too faint episodes, with respect to what sub-mm counts
indicate. To fix this problem, the simplified prescriptions used
in semi-analytical models to describe the behavior of baryons in
DM halos need some substantial revision, on which work is in
progress (Lacey et al, in preparation). It is clear that, in some
sense, semi-analytical models, based on the hierarchical clustering
paradigm, should be modified in such a way that the formation of
a subset of galaxies resemble the prediction of the so called
"monolithic scenario" (Eggen, Lynden-Bell \& Sandage, 1962;
Larson, 1975)

\begin{figure} 
\centerline{\epsfig{file=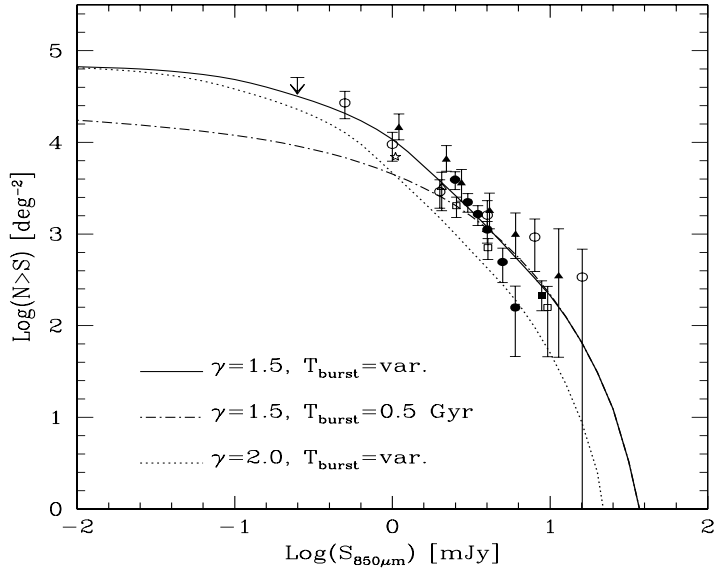,width=10truecm,height=8truecm}}
\caption{The formation rate for spheroids derived from the QSO
luminosity function evolution, coupled with plausible assumption
for their SED evolution can easily reproduce the $850\mu$m SCUBA
counts. Adapted from Granato et al.\ (2001).}
\end{figure}

Further indications of this can be gained taking advantage of the
observational evidence suggesting that QSOs did shine in the core
of early-type protogalaxies, during their main episode of star
formation. Granato et al (2001) used this information to derive
the formation rate of spheroids as a function of redshift,
essentially by means of a deconvolution of the rather well known
evolution of the QSO luminosity function. Then they used GRASIL
(Silva et al 1998) to predict their SED evolution in the FIR
sub-mm region. In this way they have shown that SCUBA counts are
well reproduced by early-type protogalaxies (Fig. 3), in
particular if the active formation phase was quicker in more
massive objects, as suggested by chemical enrichment studies
(Thomas, Greggio \& Bender 1999 and references therein).

\section{Prospects for the future}

Although mm observations have had a strong impact on models for
the formation and evolution of galaxies in the past few years,
and are leading to a major revision of them, a systematic study of the
high-z universe is almost impossible with present instruments,
which are limited by a small accessible area and poor resolution.
As a consequence, up to now relatively few objects have been
detected, and their optical identification, which is at present a
necessary step to obtain a spectroscopic redshift, is difficult.
Also, the confusion limits at faint fluxes are severe: the
confusion noise may be important at $\lesssim 2$ mJy for SCUBA
(Blain 2000).

In summary, we have now only $\sim 10^2$ sub-mm selected sources
mostly without a key information: a reliable redshift. For most
of the sources we have only "photometric" redshifts (or lower
limits), based on ratios of fluxes in two different sub-mm bands
or on sub-mm to radio ratios.

The detection rates for surveys to be performed with future IR and
submillimetric facilities (e.g.\ SIRTF, LMT, ALMA, Herschel-FIRST)
are expected to increase by orders of magnitude with respect to
present ones (e.g.\ figure 4 in Blain 1999). A fundamental point
is that these large samples could be of little use to further
constrain the history of star formation in the universe, unless
good estimates of the redshits will be available. On one hand, it
will be possible to combine information coming from surveys at
different wavelengths, if properly planned, to estimate
photometric redshifts (see also Silva et al contribution to these
proceedings). On the other hand the development of the "redshift
machine" at LMT and ALMA, based on an order of magnitude
increase of the bandwidth of sub-mm detectors, will allow to
determine the spectroscopic redshift of the sources directly from
CO transitions, without the need of the optical identification (see
Carrasco contribution).

\acknowledgements We would like to thank the organizers of this
meeting for their hospitality and for financial support. GLG has
been in part supported also by ASI contract 1/R/27/00 and by SAGG, LS
by Cofin99. We also acknowledge the contribution of our GALFORM collaborators
(Cedric Lacey, Carlton Baugh, Shaun Cole and Carlos Frenk)  to
the work presented here. Special thanks are due to our closest
collaborators Alessandro Bressan, Gigi Danese, Gianfranco De
Zotti and Pasquale Panuzzo.

\end{article}
\end{document}